\documentclass[preprint,eqsecnum,aps,showpacs]{revtex4}
\usepackage{latexsym}
\begin{document}

\newcommand{\bea}{\begin{eqnarray}}
\newcommand{\eea}{\end{eqnarray}}
\newcommand{\be}{\begin{equation}}
\newcommand{\ee}{\end{equation}}

%
%
\def\shiftleft#1{#1\llap{#1\hskip 0.04em}}
\def\shiftdown#1{#1\llap{\lower.04ex\hbox{#1}}}
\def\thick#1{\shiftdown{\shiftleft{#1}}}
\def\b#1{\thick{\hbox{$#1$}}}

\title{Spin of ground state baryons}

\author{
A. J. Buchmann$^{1}$\thanks{email:alfons.buchmann@uni-tuebingen.de}, 
E. M. Henley$^{2}$\thanks{email:henley@phys.washington.edu} \\
Institut f\"ur Theoretische Physik, Universit\"at T\"ubingen\\
Auf der Morgenstelle 14, D-72076 T\"ubingen, Germany \\
$^2$ Department of Physics and Institute for Nuclear Theory, Box 351560, \\ 
University of Washington, Seattle, WA 98195, U.S.A.\\ }


\begin{abstract}
We calculate the quark spin contribution to the total angular momentum
of flavor octet and flavor decuplet 
ground state baryons using a spin-flavor symmetry based 
parametrization method of quantum chromodynamics. We find that 
third order SU(6) symmetry breaking three-quark operators 
are necessary to explain the experimental result $\Sigma_1=0.32(10)$.
For spin 3/2 decuplet baryons we predict that the quark spin contribution 
is $\Sigma_3=3.93(22)$, i.e. considerably larger than their total angular 
momentum.
\end{abstract}

\maketitle

\section{Introduction}

The question how the proton spin is made up from 
the quark spin $\Sigma$, quark orbital angular momentum $L_q$,
gluon spin $\Delta g$, and gluon orbital angular momentum $L_g$
\be
J = \frac{1}{2} \Sigma + L_q + \Delta g + L_g 
\ee
is one of the central issues in nucleon structure physics~\cite{seh74,ji97}.
In the constituent quark model with only one-body operators
one obtains $J=\Sigma/2 =1/2$, i.e., the proton spin is the sum
of the constituent quark spins and nothing else. 
However, experimentally it is known that only about 1/3 of  the proton 
spin comes from quarks~\cite{ash89,abe98}.
The disagreement between the quark model 
result and experiment came as a surprise because the same model 
accurately described the related proton and neutron magnetic moments.

Using a broken SU(6) spin-flavor symmetry based 
parametrization of quantum chromodynamics we 
show that the failure of the  quark model to describe 
the quark contribution to proton spin correctly is due to the neglect 
of three-quark terms in the axial current. Including third order SU(6) 
symmetry breaking three-quark terms 
we reproduce the measured quark contribution to the proton spin and predict 
the spin carried by quarks for the remaining flavor octet and decuplet 
ground state baryons. 

\section{Spin-flavor symmetry and QCD parametrization method}
\label{sec:QCD}

We use a general parametrization method developed 
by Morpurgo~\cite{Mor89} to calculate the quark contribution 
to baryon spin in a systematic manner. 
This method is based on broken spin-flavor
symmetry and quark-gluon dynamics of quantum chromodynamics (QCD).
The basic idea is to {\it formally} define, for the observable at
hand, a QCD operator $\Omega$ and QCD eigenstates $\vert B \rangle$
expressed explicitly in terms of quarks and gluons. The corresponding matrix
elements can, with the help of the unitary operator $V$, be reduced to an
evaluation in a basis of pure three-quark states
$\vert\Phi_B \rangle $ with orbital angular momentum
$L=0$
\begin{equation}
\label{map}
\left \langle B \vert \Omega \vert B \right \rangle =
\left \langle \Phi_B \vert
V^{\dagger}\Omega V \vert \Phi_B \right \rangle =
\left \langle W_B \vert
{ {\tilde \Omega}} \vert W_B \right \rangle \, .
\end{equation}
The spin-flavor wave functions
contained in $\vert \Phi_B \rangle$ are denoted by $\vert W_B\rangle $.
The operator $V$ dresses the pure three-quark
states with $q\bar q$ components and gluons and
thereby generates
the exact QCD eigenstates $\vert B \rangle $.

The main task is to find the most general expression 
for the operator ${{\tilde \Omega}}$ that is
compatible with the space-time and inner QCD symmetries.
Usually, this is a sum of one-, two-, and three-quark
operators in spin-flavor space multiplied by {\it a priori} 
unknown constants which parametrize the orbital and color space
matrix elements. 
Empirically, a hierarchy in the importance
of one-, two-, and three-quark operators is found.
This fact can be understood in the $1/N_c$ expansion of QCD where
two- and three-quark operators describing second and third
order SU(6) symmetry breaking
are usually suppressed by powers of $1/N_c$ and $1/N_c^2$
respectively compared to one-quark operators associated with first order
symmetry breaking~\cite{Leb00}. 
Previously, we have applied this method 
also  to calculate higher order corrections to 
baryon-meson couplings as well as baryon 
electromagnetic moments~\cite{Hen08}.
 
To begin with, we show which spin-flavor structures 
contribute to the flavor singlet axial current. 
Generally, an SU(6) spin-flavor symmetry breaking operator 
${\tilde \Omega}^{R}$ acting on the ${\bf 56}$ dimensional baryon ground state 
supermultiplet must transform according to one of the irreducible 
representations $R$ contained in the direct product~\cite{Sak64}
\be
\label{directproduct}
\bar{{\bf 56}} \times {\bf 56}
=  {\bf 1} + {\bf 35} + {\bf 405} + {\bf 2695}.
\ee
The ${\bf 1}$ dimensional representation (rep) on the right-hand side
of Eq.(\ref{directproduct}) corresponds to an SU(6) symmetric operator, 
while the ${\bf 35}$, ${\bf 405}$,
and ${\bf 2695}$ dimensional reps characterize respectively, first, second, 
and third order SU(6) symmetry breaking. 
Therefore, a general SU(6) breaking operator has the form 
\be
\label{genop}
{\tilde \Omega} 
=  {\tilde \Omega} ^{\bf 1} + {\tilde \Omega} ^{\bf 35} + 
{\tilde \Omega} ^{\bf 405} + {\tilde \Omega} ^{\bf 2695}.
\ee
In terms of quarks, the operators on the right-hand side of 
Eq.(\ref{genop}) are represented respectively 
by zero-, one-, two-, and three-quark operators~\cite{leb95}.
The two- and three-quark operators are an effective description 
of quark-antiquark and gluon degrees of freedom 
that have been moved from the QCD eigenstates $\vert B\rangle$ 
to the operator ${\tilde \Omega}$ by virtue of the unitary 
transformation $V$. 

We now decompose each SU(6) tensor ${\tilde \Omega} ^R$ 
into subtensors ${\tilde \Omega} ^R_{(F,2J+1)}$ 
with definite transformation properties with respect to the 
SU(3)$_F$  flavor and SU(2)$_J$ spin subgroups of SU(6). 
To describe spin, flavor singlet, axial vector 
subtensors of type ${\tilde \Omega}^R_{(1,3)}$ are required. 
Denoting operators by their dimensionalities for simplicity we 
obtain~\cite{Beg64a,coo65} 
\bea 
\label{decomp}
{\bf 35}& = &
({\bf 8},{\bf 3}) +
({\bf 8},{\bf 1}) +
({\bf 1},{\bf 3}), \nonumber \\
{\bf 405}
& = &
 ({\bf 27},{\bf 5}) +
 ({\bf 8}, {\bf 5}) +
({\bf 1}, {\bf 5}) +
({\bf 27},{\bf 3}) +
({\bf 10},{\bf 3}) +  ({\bar {\bf 10}},{\bf 3}) + 2 \,  ({\bf 8},{\bf 3})
\nonumber \\
& + &
({\bf 27},{\bf 1}) +
({\bf 8},{\bf 1}) +
({\bf 1},{\bf 1}), \nonumber \\
{\bf 2695}
& = &
({\bf 64},{\bf 7}) +
({\bf 27},{\bf 7}) +
({\bf 8},{\bf 7}) + 
({\bf 1},{\bf 7}) 
\nonumber \\
& + &
({\bf 64},{\bf 5}) +
({\bf 35},{\bf 5}) +
({\bar {\bf 35}},{\bf 5}) +
2\, ({\bf 27},{\bf 5}) +
({\bf 10},{\bf 5}) + 
({\bar {\bf 10}},{\bf 5}) +
2 ({\bf 8},{\bf 5})  
\nonumber \\
& + &
({\bf 64},{\bf 3}) +
({\bf 35},{\bf 3}) +
({\bar {\bf 35}},{\bf 3}) +
3\, ({\bf 27},{\bf 3}) +
2\, ({\bf 10},{\bf 3}) + 
2 ({\bf 8},{\bf 3}) +   
({\bf 1},{\bf 3})
\nonumber \\
& + &
({\bf 64},{\bf 1}) +
({\bf 27},{\bf 1}) +
({\bf 10},{\bf 1}) + 
({\bar {\bf 10}},{\bf 1}) +
({\bf 8},{\bf 1}).
\eea
Here, the first and second entries in the parentheses refer to the dimensions
of the SU(3)$_F$ and SU(2)$_J$ reps according to which the corresponding
operators transform.

From the flavor-spin decompositions in 
Eq.(\ref{decomp}) it is clear that a flavor singlet axial vector operator
(${\bf 1}, {\bf 3})$ needed to describe baryon spin, is contained only in the  
${\bf 35}$ and ${\bf 2695}$ dimensional reps of SU(6) so that we can write
${\bf A} : = {\tilde \Omega} =  
{\tilde \Omega}^{35}_{({\bf 1},{\bf 3})}+
{\tilde \Omega}^{2695}_{({\bf 1},{\bf 3})}$.  
As a result, the flavor singlet axial vector 
current contains only a one-quark and a three-quark term
\begin{equation}
\label{total} 
{\bf A} = {\bf {A}}_{[1]} + {\bf {A}}_{[3]}  
= A \, \sum_{i=1}^3 \  {\b{\sigma}}_{i} + 
C \, \sum_{i \ne j \ne k}^3  \ {\b{\sigma}}_i \cdot {\b{\sigma}}_j
\ {\b{\sigma}}_{k},
\end{equation} 
where $\b{\sigma}_i$ is the Pauli spin matrix of quark $i$ and the constants 
$A$ and $C$ are to be determined from experiment.
Two-quark operators, such as 
$\sum_{i\ne j} {\b{\sigma}}_i \times {\b{\sigma}}_j$ or 
$\sum_{i\ne j} ({\b{\sigma}}_i \cdot {\b{\sigma}}_j) \, 
{\b{\sigma}}_i $ 
add up to zero or can be reduced to one-body operators.
Therefore, there is no two-quark contribution to the
flavor singlet axial vector current ${\bf A}$, 
in agreement with the general symmetry argument that 
the ${\bf 405}$ rep does not contain a (${\bf 1}, {\bf 3})$ 
operator structure. 

In Ref.~\cite{bar06} a flavor singlet 
two-body gluon exchange current ${\bf A}_0$ was constructed 
from the flavor octet axial axial current ${\bf A}_8$ 
by replacing the Gell-Mann matrix $\b{\lambda}_8$ with the SU(3) flavor
singlet matrix $\b{\lambda}_0$.
From the perspective of broken SU(6) symmetry, such a {\it two}-body 
exchange current, which has also been used in Ref.~\cite{myh08}, 
does not exist. 

\section{Spin-flavor wave functions} 

The SU(6) spin-flavor wave functions 
of octet baryons are, for example, for the proton in standard notation  
\begin{eqnarray}
\label{wf1}
\vert p \uparrow \rangle  =  \frac{1}{\sqrt{2}} & & 
\biggl \lbrace \frac{1}{\sqrt{6}}
\biggl \vert \left ( 2uud - udu -duu \right ) \biggr \rangle \ 
\frac{1}{\sqrt{6}} \biggl \vert 
\left ( 2 \uparrow \uparrow \downarrow - \uparrow \downarrow \uparrow
- \downarrow \uparrow \uparrow \right ) \biggr \rangle \nonumber \\
& & + \frac{1}{\sqrt{2}} \biggl \vert 
\left ( udu -duu \right ) \biggr \rangle  \ 
\frac{1}{\sqrt{2}} \biggl \vert 
\left ( \uparrow \downarrow \uparrow - \downarrow \uparrow \uparrow \right )
\biggr \rangle  
\biggr \rbrace .
\end{eqnarray}
Alternatively, one can write this wave function as
\begin{eqnarray}
\label{wf2}
\vert p \uparrow \rangle  =  \frac{1}{\sqrt{18}} & & 
\biggl \vert 
     \phantom{+} 2u \uparrow   u \uparrow    d \downarrow 
     -\, u \uparrow   u \downarrow  d \uparrow 
     -\, u \downarrow u \uparrow    d \uparrow  \nonumber \\
& & +\, 2d \downarrow u \uparrow    u \uparrow  
     -\, d \uparrow   u \downarrow  u \uparrow  
     -\, d \uparrow   u \uparrow    u \downarrow \nonumber \\ 
& & +\, 2u \uparrow   d \downarrow  u \uparrow   
     -\, u \uparrow   d \uparrow    u \downarrow 
     -\, u \downarrow d \uparrow    u \uparrow 
\biggr \rangle.
\end{eqnarray}
Likewise one can write down the spin-flavor wave functions
of other octet and decuplet baryons~\cite{lic70}.

\section{Quark spin contribution to baryon spin}

The matrix elements of the quark contribution to baryon spin  
can be straightforwardly calculated by sandwiching the 
flavor singlet axial current ${\bf A}$ of Eq.(\ref{total})
between SU(6) baryon wave functions, e.g. for the proton in Eq.(\ref{wf2}). 
Our results for the spin 1/2 octet and the spin 3/2 decuplet baryons 
are 
\bea
\label{matrixelements} 
\Sigma_{1}: & = &
\langle B_8 \uparrow \vert {\bf A}_z \vert  B_8 \uparrow \rangle = A - 10\, C 
\nonumber \\
\Sigma_{3}: & = &
\langle B_{10} \uparrow \vert {\bf A}_z \vert  B_{10} \uparrow \rangle = 
3\,A + 6\,C, 
\eea
where $B_8$ ($B_{10}$) stands for any member of the 
baryon flavor octet (decuplet). Here, $\Sigma_i$ is twice the quark spin
contribution to the total baryon angular momentum.
We predict the same quark contribution to the total baryon 
angular momentum for all members 
of a given flavor multiplet independent of the flavor composition 
of individual baryons. This is to be expected because 
we are dealing with a flavor singlet operator that does not break
SU(3) symmetry. On the other hand, we find that SU(6) symmetry 
is broken as reflected by the different expressions for 
flavor octet and decuplet baryons.

To calculate the contribution of individual quark flavors 
we define one-body $u$-quark and $d$-quark operators 
acting only on $u$-quarks and $d$-quarks as
\be
\label{u-quark1b}
{\bf A}_{[1]\, z}^u = A \sum_{i=1}^{3} \b{\sigma}^u_{i\, z}, \qquad
{\bf A}_{[1]\, z}^d = A \sum_{i=1}^{3} \b{\sigma}^d_{i\, z},
\ee
with matrix elements between proton wave functions
\be
\langle p \uparrow \vert A\, \sum_{i=1}^3\  
\b{\sigma}^u_{i\, z} \vert p \uparrow \rangle  = \frac{4}{3} A, \qquad
\langle p \uparrow \vert A\, \sum_{i=1}^3\ 
\b{\sigma}^d_{i\, z} \vert p \uparrow \rangle = -\frac{1}{3} A.
\ee
Adding both contributions gives 
\be
\langle p \uparrow \vert A\, \sum_{i=1}^3\ 
\left ( \b{\sigma}^u_{i\, z} + \b{\sigma}^d_{i\, z} \right )
\vert p \uparrow \rangle 
= \frac{4}{3} A - \frac{1}{3} A = A,
\ee
in agreement with the first term of Eq.(\ref{matrixelements}).

Analogously, three-body $u$-quark and $d$-quark operators are defined as 
\bea
\label{u-quark3b} 
{\bf {A}}_{[3]\, z}^u & = &
2 C  \biggl ( 
 {\b{\sigma}}^u_1 \cdot {\b{\sigma}}^d_2\ {\b{\sigma}}^u_{3\, z} 
+ \, {\b{\sigma}}^d_1 \cdot {\b{\sigma}}^u_2\ {\b{\sigma}}^u_{3\, z} 
+ {\b{\sigma}}^u_1 \cdot {\b{\sigma}}^d_3\ {\b{\sigma}}^u_{2\, z}  
+ \,  {\b{\sigma}}^d_1 \cdot {\b{\sigma}}^u_3\ {\b{\sigma}}^u_{2\, z}
+ {\b{\sigma}}^u_2 \cdot {\b{\sigma}}^d_3\ {\b{\sigma}}^u_{1\, z}
+ \, {\b{\sigma}}^d_2 \cdot {\b{\sigma}}^u_3\ {\b{\sigma}}^u_{1\, z}  \biggr ),
\nonumber \\
{\bf {A}}_{[3]\, z}^d & = & 
C  \biggl (  
 {\b{\sigma}}^u_1 \cdot {\b{\sigma}}^u_2\ {\b{\sigma}}^d_{3\, z} 
+ \, {\b{\sigma}}^u_2 \cdot {\b{\sigma}}^u_1\ {\b{\sigma}}^d_{3\, z} 
+  {\b{\sigma}}^u_1 \cdot {\b{\sigma}}^u_3\ {\b{\sigma}}^d_{2\, z}  
+ \,  {\b{\sigma}}^u_3 \cdot {\b{\sigma}}^u_1\ {\b{\sigma}}^d_{2\, z}
+   {\b{\sigma}}^u_2 \cdot {\b{\sigma}}^u_3\ {\b{\sigma}}^d_{1\, z}
+ \, {\b{\sigma}}^u_3 \cdot {\b{\sigma}}^u_2\ {\b{\sigma}}^d_{1\, z}\biggr ),
\nonumber \\
\eea
with matrix elements 
\be
\langle p \uparrow \vert \, {\bf A}_{[3]\, z}^u \vert 
p \uparrow \rangle = -\frac{28}{3} \, C, \qquad
\langle p \uparrow \vert \, {\bf A}_{[3]\, z}^d \vert 
p \uparrow \rangle = - \frac{2}{3}\, C. 
\ee
The total three-quark spin contribution to proton spin is
\be
\langle p \uparrow \vert \, 
{\bf A}_{[3]\, z}^u + {\bf A}_{[3]\, z}^d 
\vert  p \uparrow \rangle = 
 -\frac{28}{3} \, C - \frac{2}{3}\, C = -10 \, C,
\ee
as it should be according to Eq.(\ref{matrixelements}).
Summarizing, we obtain for the $u$- and $d$-quark contributions
to the spin of the proton
\bea
\label{flavordecomp}
\Delta u: = 
\langle p \uparrow \vert \, 
{\bf A}_{[1]\, z}^u + {\bf A}_{[3]\, z}^u  \vert  p \uparrow \rangle 
& = & \phantom{-}\frac{4}{3}\, A  - \frac{28}{3} \, C 
\nonumber \\
\Delta d : =
\langle p \uparrow \vert \, 
{\bf A}_{[1]\, z}^d + {\bf A}_{[3]\, z}^d  \vert  p \uparrow 
\rangle 
& = & -\frac{1}{3}\, A  - \frac{2}{3} \, C. 
\eea

We fix the constants $A$ and $C$ as follows.
The combined deep inelastic scattering and hyperon $\beta$-decay data
give~\cite{abe98}
\begin{eqnarray}
\label{deltaq}
\Delta u = \hspace{.35cm} 0.83 \pm 0.03,\hspace{.5cm}
\Delta d =-0.43 \pm 0.03, \hspace{.5cm}
\Delta s = -0.09 \pm 0.04.
\end{eqnarray}
The sum of these experimental spin fractions
$\Sigma_{1_{exp}}
=\Delta u + \Delta d + \Delta s = 0.32(10)$ is considerably smaller
than expected from the additive quark model, which gives 
$\Sigma_1=1$. Solving Eq.(\ref{flavordecomp}) for $A$ and $C$ we get 
\bea
A & = & \phantom{-}\frac{1}{6}\, \, \Delta u  - \frac{7}{3}\, \Delta d,  
\nonumber \\
C & = & -\frac{1}{12}\, \Delta u - \frac{1}{3}\, \Delta d. 
 \eea
Inserting the experimental results for $\Delta u$ and $\Delta d$ 
from Eq.(\ref{deltaq}) we obtain $A=1.15(7)$ and $C=0.08(1)$.  

According to Eq.(\ref{matrixelements}) the quark spin contribution 
to the total angular momentum of flavor octet baryons 
is then $\Sigma_1 = 0.35(12)$ compared to the experimental result 
$\Sigma_{1_{exp}}= 0.32(10)$.
Despite the typical $1/N_c^2 \cong 1/9$ suppression of three-quark 
compared to one-quark terms, for octet baryon spin,  
the three-quark contribution is of the same importance 
as the one-quark term because of the factor 10 multiplying the $C$ term.
As a result of the cancellation between the one-quark and the three-quark
terms one finds that quark spin contributes only 1/3 of the total 
octet baryon spin. As mentioned before, the three-quark term 
is an effective description of quark-antiquark and gluon degrees of freedom
in physical baryons. These degrees of freedom are also responsible for 
orbital angular momentum that is needed to obtain the total baryon spin.

With $A$ and $C$ fixed, we predict the quark spin contribution 
to decuplet baryon spin not considered before by other authors as
\be
\Sigma_{3}  =
\langle B_{10} \uparrow \vert {\bf A} \vert  B_{10} \uparrow \rangle = 
3\,A + 6\,C = 3.93(22).  
\ee
It is interesting that for decuplet baryons, quark spins 
add up to 1.31 times the additive quark model value $2 S_z = 3$. 
Therefore, while orbital angular momentum must provide a 
positive contribution for octet baryons, it must reduce the quark spin
in the case of decuplet baryons. 

\section{Summary}

In summary, we have presented a straightforward calculation of the 
quark spin contribution to the total angular momentum of ground state
baryons using a spin-flavor symmetry based parametrization
of QCD.  For flavor octet baryons, we have shown that three-quark 
operators modify the standard quark model prediction based on 
one-quark operators from $\Sigma_1 =1$ to 
$\Sigma_1 = 0.35(12)$ in agreement with the experimental result.  
On the other hand, in the case of flavor decuplet baryons, three-quark 
operators enhance the contribution of one quark operators from 
$\Sigma_3=3$ to $\Sigma_3=3.93(22)$. 

In  this paper, our concern has been with the quark spin contribution 
to baryon total angular momentum.
There are two further contributions, the orbital angular momentum 
of the quarks and the gluon contributions. 
Concerning gluon spin $\Delta g$, different experiments 
agree that it is too small to explain 
the total proton angular momentum~\cite{tho09}. Furthermore, it 
has been shown that gluon orbital angular momentum $L_g$ 
is very small as well~\cite{bro06}.
Quark orbital angular momentum $L_q$ has recently been investigated by 
a number of authors~\cite{tho08,wak08,gar10}. 
We do not attempt to discuss these contributions here, because 
it would detract from our main theme. We hope to address this issue in 
a future publication.

\noindent
{\bf Acknowledgments:} 
One of the authors (EMH) is grateful to the Humboldt Foundation 
for a grant and to Wolfram Weise and the Technical University of Munich 
for their hospitality.

\end{document}